\RequirePackage{fix-cm}
\RequirePackage{units}
\documentclass[twocolumn,epjc3]{svjour3}                     
\smartqed  
\usepackage{graphicx}
\usepackage{amssymb}
\usepackage{amsmath}
\usepackage{lineno}
\usepackage{color}
\usepackage{hyperref}
\usepackage{caption}
\usepackage{subcaption}
\usepackage{siunitx} 
\usepackage{booktabs}

\usepackage{placeins}
\usepackage{microtype}

\newcommand{\numu}{$\nu_{\mu}$ }

\newcommand{\dg}{^\circ}
\newcommand{\truee}{\textsc{Truee}}

\begin{document}

\title{Measurement of the $\nu_{\mu}$ energy spectrum with IceCube-79
}
\subtitle{}





\onecolumn
\author{IceCube Collaboration: M.~G.~Aartsen\thanksref{Adelaide}
\and M.~Ackermann\thanksref{Zeuthen}
\and J.~Adams\thanksref{Christchurch}
\and J.~A.~Aguilar\thanksref{BrusselsLibre}
\and M.~Ahlers\thanksref{MadisonPAC}
\and M.~Ahrens\thanksref{StockholmOKC}
\and I.~Al~Samarai\thanksref{Geneva}
\and D.~Altmann\thanksref{Erlangen}
\and K.~Andeen\thanksref{Marquette}
\and T.~Anderson\thanksref{PennPhys}
% ...
}
\authorrunning{IceCube Collaboration}
\thankstext{a}{Earthquake Research Institute, University of Tokyo, Bunkyo, Tokyo 113-0032, Japan}
\institute{
% ...
Department of Physics, University of Adelaide, Adelaide, 5005, Australia \label{Adelaide}
% ...
\and Universit\'e Libre de Bruxelles, Science Faculty CP230, B-1050 Brussels, Belgium \label{BrusselsLibre}
% ...
\and Dept.~of Physics and Astronomy, University of Canterbury, Private Bag 4800, Christchurch, New Zealand \label{Christchurch}
% ...
\and Erlangen Centre for Astroparticle Physics, Friedrich-Alexander-Universit\"at Erlangen-N\"urnberg, D-91058 Erlangen, Germany \label{Erlangen}
\and D\'epartement de physique nucl\'eaire et corpusculaire, Universit\'e de Gen\`eve, CH-1211 Gen\`eve, Switzerland \label{Geneva}
% ...
\and Dept.~of Physics and Wisconsin IceCube Particle Astrophysics Center, University of Wisconsin, Madison, WI 53706, USA \label{MadisonPAC}
% ...
\and Department of Physics, Marquette University, Milwaukee, WI, 53201, USA \label{Marquette}
% ...
\and Oskar Klein Centre and Dept.~of Physics, Stockholm University, SE-10691 Stockholm, Sweden \label{StockholmOKC}
% ...
\and Dept.~of Physics, Pennsylvania State University, University Park, PA 16802, USA \label{PennPhys}
% ...
\and DESY, D-15735 Zeuthen, Germany \label{Zeuthen}
}
\date{Received: date / Accepted: date}
\maketitle
\twocolumn


\begin{abstract}

IceCube is a neutrino observatory deployed in the glacial ice at the geographic South Pole. The $\nu_\mu$ energy unfolding described in this paper is based on data taken with IceCube in its 79-string configuration.

A sample of muon neutrino charged-current interactions with a purity of 99.5\% was selected by means of a multivariate classification process based on machine learning.
The subsequent unfolding was performed using the software \truee. The resulting spectrum covers an E$_\nu$-range of more than four orders of magnitude from 125 GeV to 3.2 PeV. Compared to the Honda atmospheric neutrino flux model, the energy spectrum shows an excess of more than $1.9\,\sigma$ in four adjacent bins for neutrino energies $E_\nu\geq177.8\,\unit{TeV}$. The obtained spectrum is fully compatible with previous measurements of the atmospheric neutrino flux and recent IceCube measurements of a flux of high-energy astrophysical neutrinos.

\end{abstract}


\section{Introduction}
\label{sec:Introduction}

The neutrino flux, which can be observed with instruments such as IceCube, has its origin both in cosmic ray air showers (atmospheric neutrinos)~\cite{amanda,IC40Atmospheric,IC59} and extraterrestrial sources~\cite{doi:10.1146/annurev.nucl.50.1.679,HESE,ic79weaver,globalfit}. These different components can be modeled separately and fitted to data, with the atmospheric component dominating up to energies of approximately $300\,\unit{TeV}$.
Such fits, however, require assumptions on the spectral shape. An extraction of the energy spectrum from experimental data is more model-independent, as no assumption on the cosmic ray composition or spectral shape is required. It thus poses an alternative to fitting model parameters and allows a direct comparison to theoretical model predictions. 

This paper presents a measurement of the muon neutrino energy spectrum with IceCube during its deployment phase in the 79-string configuration (IC79). The spectrum was obtained from a highly pure sample of neutrino candidates by means of regularized unfolding.

\subsection{The IceCube Detector}
\label{subsec:IceCube}

The IceCube detector, located at the geographic South Pole, is a neutrino observatory with an instrumented volume of one cubic kilometer~\cite{instrumentation}. It consists of $5160$ digital optical modules (DOMs) deployed on $86$ strings at depths between $1450\,\unit{m}$ and $2450\,\unit{m}$. The strings are arranged in a hexagonal array, with a string-to-string distance of $125\,\unit{m}$. The $86$ strings include the low-energy extension DeepCore~\cite{deepcore}, which has a string spacing of approximately $70\,\unit{m}$ and a vertical DOM distance of $7~\unit{m}$. It is optimized for low energies and reduces the energy threshold of the entire detector to $E_{\text{th}}\sim 10\,\unit{GeV}$~\cite{deepcore}. 

Each DOM consists of a glass sphere of $35.6\,\unit{cm}$ diameter, which houses a $25~\unit{cm}$ Hamamatsu R7081-02 photomultiplier tube (PMT) and a suite of electronics board assemblies. Internal digitizing and time-stamping the photonic signals ensures high accuracy and a wide dynamic range of the DOMs. Packaged digitized data is then
transmitted to the IceCube Laboratory (ICL) at the South Pole. Each DOM can operate as a complete and autonomous data acquisition system~\cite{instrumentation,IceCubeDAQ}. The air shower array IceTop complements the detector~\cite{icetop}. 

As neutrinos cannot be observed directly, they are detected via secondary particles produced in the interactions of neutrinos with nuclei in the ice or the bedrock. These secondary particles induce the emission of Cherenkov light, which is recorded by the DOMs. The majority of the events observed with IceCube are \emph{track-like} events, which originate from muons propagating through the detector. These muons are either produced in charged current (CC) neutrino-nucleon interactions or in cosmic ray air showers. The second most frequent signature are \emph{cascade-like} events, originating from CC interactions of $\nu_{e}$ and $\nu_{\tau}$, where the second cascade, resulting from the decay of the emerging $\tau$-lepton, cannot be experiementally resolved. \emph{Cascade-like} events further originate from neutral-current (NC) interactions of neutrinos of all flavors within the instrumented volume.

\subsection{Atmospheric Muon Neutrinos}
\label{subsec:AtmosphericNeutrinos}

The atmospheric muon neutrino flux is expected to consist of two components distinguished by the lifetime of their hadronic parent particles. Conventional atmospheric muon neutrinos originate from the decay of charged pions and kaons in cosmic ray air showers. Due to their relatively long lifetime ($\tau \sim 10^{-8}\,\unit{s}$~\cite{PDG}), kaons and pions interact prior to decaying. This results in a flux of approximately $\frac{\mathrm{d}\Phi}{\mathrm{d}E} \propto E^{-3.7}$. 

The second, much rarer component, consisting of prompt atmospheric neutrinos, originates from the decay of charmed particles such as $D$ mesons and $\Lambda_{\text{c}}^+$ baryons. Due to their short lifetime ($\tau \sim 10^{-12}\,\unit{s}$~\cite{PDG}), these hadrons decay before interacting. Prompt atmospheric neutrinos inherit the spectral index of the cosmic ray flux directly, resulting in a flux of $\frac{\mathrm{d}\Phi}{\mathrm{d}E} \propto E^{-2.7}$. The conventional component is the dominant component to the flux of atmospheric neutrinos up to energies of $E_{\nu} \sim 300\,\unit{TeV}$~\cite{Enberg}. The prompt component has not been observed so far and the exact threshold depends strongly on the underlying theoretical model.

The atmospheric \numu energy spectrum has been measured by various experiments, including AMANDA~\cite{amanda}, Fr\'{e}jus~\cite{frejus} and ANTARES~\cite{antares}, as well as IceCube in the 40-~\cite{IC40Atmospheric} and 59-string configuration~\cite{IC59}.

An additional contribution to the overall flux is expected to arise from a flux of astrophysical neutrinos~\cite{Becker2008}. This flux has recently been discovered by IceCube~\cite{HESE,HESE2}. Its sources are still unknown. The spectral index of the astrophysical component is expected to be approximately $\gamma=2.0$ for the simplest assumption of Fermi acceleration~\cite{doi:10.1146/annurev.nucl.50.1.679}. Recent measurements by IceCube obtained indices between $\gamma=2.13\pm0.13$~\cite{Aachen6yrs} and $\gamma=2.50^{+0.08}_{-0.09}$~\cite{globalfit}.

A major challenge in the measurement of muon neutrinos is the background of atmospheric muons. Although muons and muon neutrinos are produced at approximately the same rate, the rate of triggering atmospheric muons is $\sim10^6$ times higher, due to the small cross sections of neutrino-nucleon interactions. 

The application of a Random Forest-based analysis chain for IC59 presented in~\cite{IC59} resulted in a high statistics sample of 66,865 atmospheric neutrino candidates with an estimated background contribution of $330\pm200$ background events. Averaging these numbers, one obtains an event rate of $9.3 \cdot 10^{-4}$ neutrino candidates per second at an average background event rate of $(3.8\pm3.4) \cdot 10^{-6}$ events per second. The separation process presented here is based on the approach presented in \cite{IC59}. Compared to~\cite{IC59}, however, the signal efficiency with respect to the starting level of the analysis was improved from 18.2\% to 26.5\% at an equally high purity of the final sample of event candidates. 

The subsequent unfolding extends the upper end of the muon neutrino energy spectrum by a factor larger than 3 (up to $3.2\,\unit{PeV}$) in comparison to previous measurements~\cite{IC59}. The obtained spectrum is in good agreement with previous measurements of the atmospheric neutrino flux up to energies of $E_\nu \approx 130\,\unit{TeV}$. For higher energies an excess above an atmospheric only assumption is observed. This excess is consistent with the flux of astrophysical neutrinos observed in other IceCube analyses and can therefore be attributed to a flux of high energy astrophysical neutrinos from unknown sources. 

The paper is organized as follows: in the next section the selection of neutrino candidates is presented. Section~\ref{sec:Unfolding} describes the unfolding of the \numu energy spectrum and discusses statistical and systematic uncertainties. A discussion of the results is given in Sec.~\ref{sec:Discussion}. The paper concludes in Sec.~\ref{sec:Conclusion} with a summary and an outlook.


\section{Event Selection}
\label{sec:EventSelection}
In this paper, data taken between the 31st of May 2010 and the 13th of May 2011 are analyzed. After data quality selections, a dataset with a livetime of 319.6 days remains for the analysis. First reconstruction and selection steps are performed at the detection site. Further processing of the data, such as detailed track reconstruction~\cite{amandaMUON} and energy estimation~\cite{energyreco}, are carried out offline. This analysis used as input a data set consisting of $\sim 2.58\times 10^{8}$ event candidates, $\sim 2.39\times10^{5}$ of which are expected to be neutrinos of atmospheric origin. One thus obtains a signal-to-background ratio of $\approx 0.93\times10^{-3}$. 

At this level of the analysis, the majority of the event candidates still consists of atmospheric muons, which need to be efficiently rejected. The separation process can be structured in two parts: straight cuts and the application of machine learning algorithms.

Simulated events provide the basis for the machine learning-based part of the event selection as well as for the applied cuts. 

Simulated neutrino events were produced using the IceCube neutrino generator NuGen, which is based on updated cross-sections for deep inelastic scattering using the HERA1.5 set of parton distribution functions~\cite{Cooper-Sakar2011}. The events were simulated according to an assumed cosmic $E^{-2}$-spectrum and weighted to neutrino flux models by Honda 2006~\cite{Honda2006} and Enberg~\cite{Enberg} to account for the conventional and the prompt component of the atmospheric muon neutrino flux. In total, $\sim8.3\times10^{6}$ simulated muon neutrino events were available at the starting level of the analysis. Since the dominant contribution to the neutrino energy spectrum arises from conventional neutrinos up to energies of $E_{\nu}\approx300\,\unit{TeV}$, the event selection does not depend on the detailed modeling of the prompt component. 

The background of atmospheric muons was simulated using the air shower code CORSIKA~\cite{CORSIKA}. The poly-gonato model~\cite{Polygonato2003} was used as an input spectrum for primary cosmic rays. In total, $\sim1.6\times10^{6}$ simulated atmospheric muons were available at the starting level of the analysis. This corresponds to approximately 6 days of detector livetime. This shortage of simulated background events was compensated by evaluating the machine learning part of the event selection in a bootstrapping procedure (see Sec.~\ref{sec:MachineLearning} for details). 

\subsection{Cuts}
\begin{figure}
\includegraphics[width=\columnwidth]{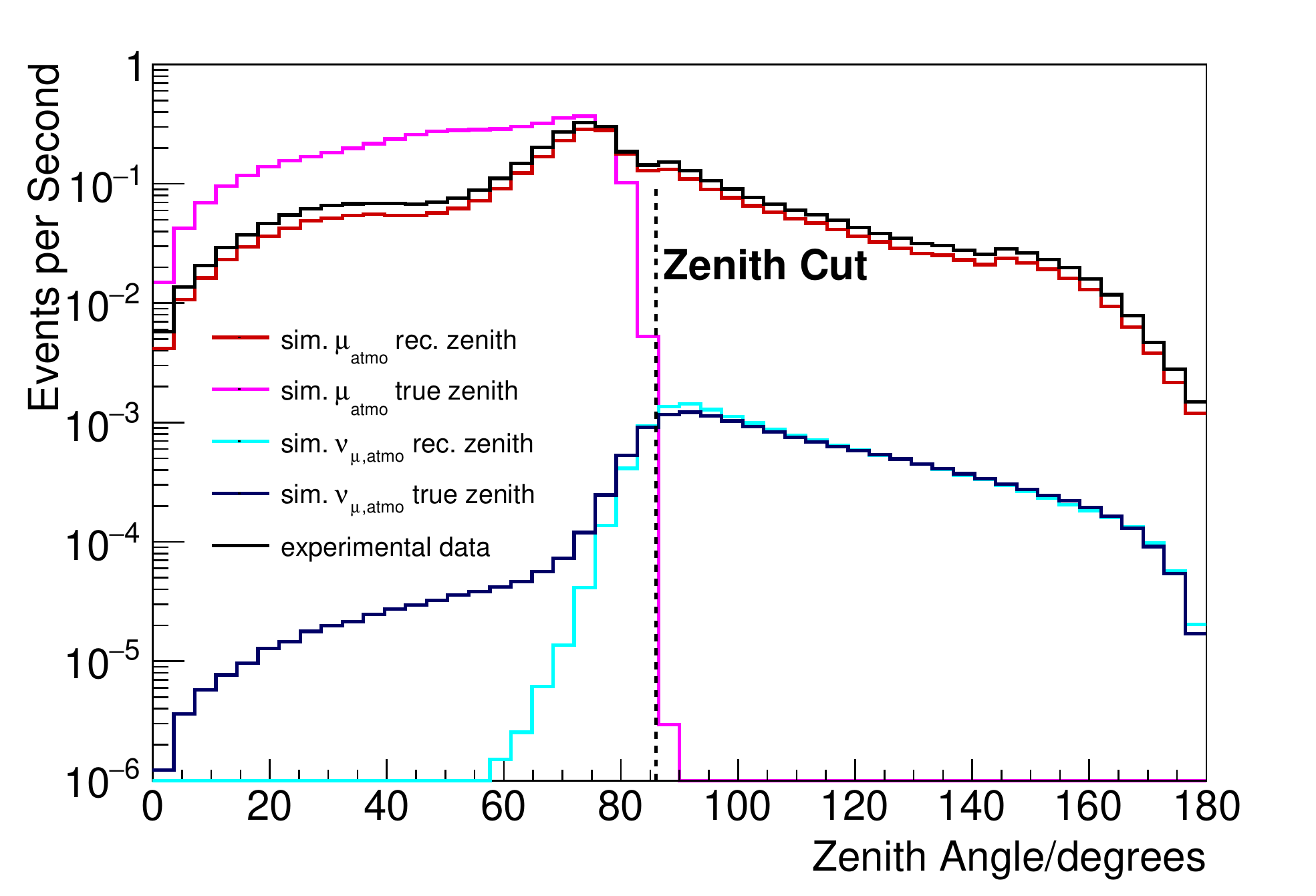}
\caption{Distributions of simulated and reconstructed zenith angles for atmospheric muons~\cite{CORSIKA}  and atmospheric muon neutrinos~\cite{ANIS}. The black dashed line indicates the position of the zenith cut applied in the separation process.}
\label{fig:truezen}
\end{figure}

Since there is no topological difference between neutrino-induced muons entering the detector and muons from cosmic ray air showers, one has two options: either to select neutrino interactions inside the instrumented volume or to select only neutrino candidates from zenith angles, for which atmospheric muons are stopped by the Earth. This analysis pursues the second approach and, in a first cut, downgoing events (zenith angle $\theta\leq86^\circ$) are rejected. The remaining background consists mostly of misreconstructed atmospheric muons, where the separation task is to distinguish between well- and misreconstructed events (see \autoref{fig:truezen}).

A second cut is applied on the reconstructed event velocity $v_{\mathrm{reco}}\geq 0.1\,\unit{c}$, which is obtained from reconstructing the muon track on the basis of the positions $\vec{r}_{i}$, and hit times $t_i$ of all DOMs giving signals in the event. Ignoring the geometry of the Cherenkov cone and the optical properties of the ice, the reconstruction fits a straight line, parameterized by the time, to the hits. The event velocity  $v_{\mathrm{reco}}$ is assumed to be constant and obtained from minimizing the following $\chi^{2}$:
\begin{equation}
\chi^{2} = \sum_{i}^{N} (\vec{r}_{i} - \vec{r}_{\textrm{LineFit}} - \vec{v}_{\textrm{LineFit}} \cdot t_{i})^{2}.
\label{eq:Chi2}
\end{equation}

The velocity is expected to be significantly smaller for \emph{cascade-like} events in comparison to high quality \emph{track-like} events. This cut therefore selects high quality \emph{track-like} events, which are required for a reliable reconstruction of the neutrino spectrum in the subsequent unfolding. Furthermore, it reduces the rate of electron neutrinos in the sample. 

A third cut is applied on the length of the reconstructed track in the detector $L_\mathrm{reco}$, which is required to be longer than $200\,\mathrm{m}$. This cut suppresses events of low energies ($E_{\nu}\leq100\,\unit{GeV}$) and events that pass near the edge of the detector. The cut favors long tracks, which are reconstructed more accurately. 

In total, the three cuts  achieve a background rejection of $81\,\%$ and keep $71\,\%$ of the neutrino-induced muons with respect to the starting level of the analysis. The cuts further favor well-reconstructed and highly energetic events. The remaining background is significantly harder to reject.

\subsection{Machine Learning}
\label{sec:MachineLearning}

The separation process is continued with algorithms from the field of machine learning and artificial intelligence. The core of the next steps is a multi-variate classification using a Random Forest (RF)~\cite{randomforest}. From the machine learning point of view, this corresponds to a two-class classification process, with neutrino-induced muons being the positive class and atmospheric muons being the negative class. To carry out the machine learning part of the analysis, a class variable was assigned to every simulated event. This class variable was chosen to be 1 for neutrino-induced muons and 0 for atmospheric muons.

As in \cite{IC59}, the RF is preceded by a variable selection and the entire classification process is embedded in multifold validation methods. This allows to control the stability and the optimization of each step of the process separately. Thereafter, recorded events are either classified as neutrino-induced muons or as misreconstructed atmospheric muon events. 

To maintain computational feasibility, not all available variables can be used as input to the Random Forest. Following the approach in~\cite{IC59}, the input variables are selected using the Minimum Redundancy Maximum Relevance~\cite{mrmr} (MRMR) algorithm. To avoid mismatches between experimental data and simulation, variables with a large $\chi^{2}$-disagreement between data and Monte Carlo were discarded from the set of candidate variables. The set of possible input variables was further reduced by removing constant variables and variables with a Pearson correlation $|\rho|\geq 0.95$. If two variables showed a correlation $|\rho|\geq 0.95$, only the first one was kept. The MRMR algorithm was then applied to this set of preselected variables. Details on the utilized implementation are presented in~\cite{Schowe/Morik/2011b}.

MRMR builds up a variable set in a sequential way. It starts with the variable with the highest correlation to the class variable. In the succeeding iterations, the $k$-th variable $V_k$ ($k>1$) is selected by taking into account the correlation $K$ of $V_k$ to the class variable (relevance), as well as the average correlation $L$ of $V_k$ to all variables $V_1, ..., V_{k-1}$ selected in the preceeding iterations (redundancy). The variable with the largest difference $D=K-L$ is added to the set. The relevance of a variable with respect to the class variable is determined by an F-test, whereas the redundancy between two variables is computed as the absolute value of the Pearson correlation coefficient~\cite{Schowe/Morik/2011b}.
This way a set of $m$ variables is built up. A more detailed description of the approach can be found in \cite{IC59} and~\cite{mrmr}.

In this analysis, $m=25$ showed a reasonable trade-off between computational feasibility and retaining information in the dataset. The selected variables can be ordered into three different groups: variables to approximate the energy, variables containing geometric properties of the event and variables indicating the reconstruction quality. Since the performance of the Random Forest depends on the agreement between data and simulation, the 25 variables selected by MRMR were manually inspected for disagreement between data and Monte Carlo. No such disagreement was found and the 25 variables were used to train the Random Forest accordingly. 

A Random Forest is an ensemble of decision trees. It is trained with simulated events to build a model that can be applied to unclassified events. In the application the $j$-th tree assigns a label $x_{i,j}=\{ 0,1 \}$ to to the $i$-th event. Thus, the final classification is achieved by averaging the output of all decision trees in the forest:
 \begin{equation}
  c_{\mathsf{Signal},i} = \frac{1}{N_{\mathsf{trees}}}\sum_j^{N_{\mathsf{trees}}} x_{i,j}.
  \label{Eq:confidence}
 \end{equation}
In machine learning, $c_{\mathsf{Signal},i}$ is generally referred to as confidence. To achieve unique trees in the RF, each decision tree is trained on a subset of simulated events. At each node only $k$ randomly chosen variables are used to find the best cut. Before applying the RF to experimental data, the RF is applied to simulated events to evaluate the performance of the classification.

After the application of the forest, the vast majority of the simulated background muons (more than 99.9\%) is found to be scored with a confidence $c_{\mathsf{Signal},i}<0.8$. Only 26 simulated atmospheric muons were found to populate the high confidence region ($c_{\mathsf{Signal}}>0.8$). Since the analysis relies on a high purity sample of neutrino candidates, the number of remaining background events needs to be estimated as accurately as possible. The confidence distribution is the basis for this estimation and thus has to be obtained as accurately as possible, as well. Due to the few background events found for $c_{\mathsf{Signal},i}\geq0.8$ the accuracy of the confidence distribution is statistically limited for this very region. This limitation can be overcome by utilizing a bootstrapping technique~\cite{bootstrap}.

\begin{figure}
\includegraphics[width=0.5\textwidth]{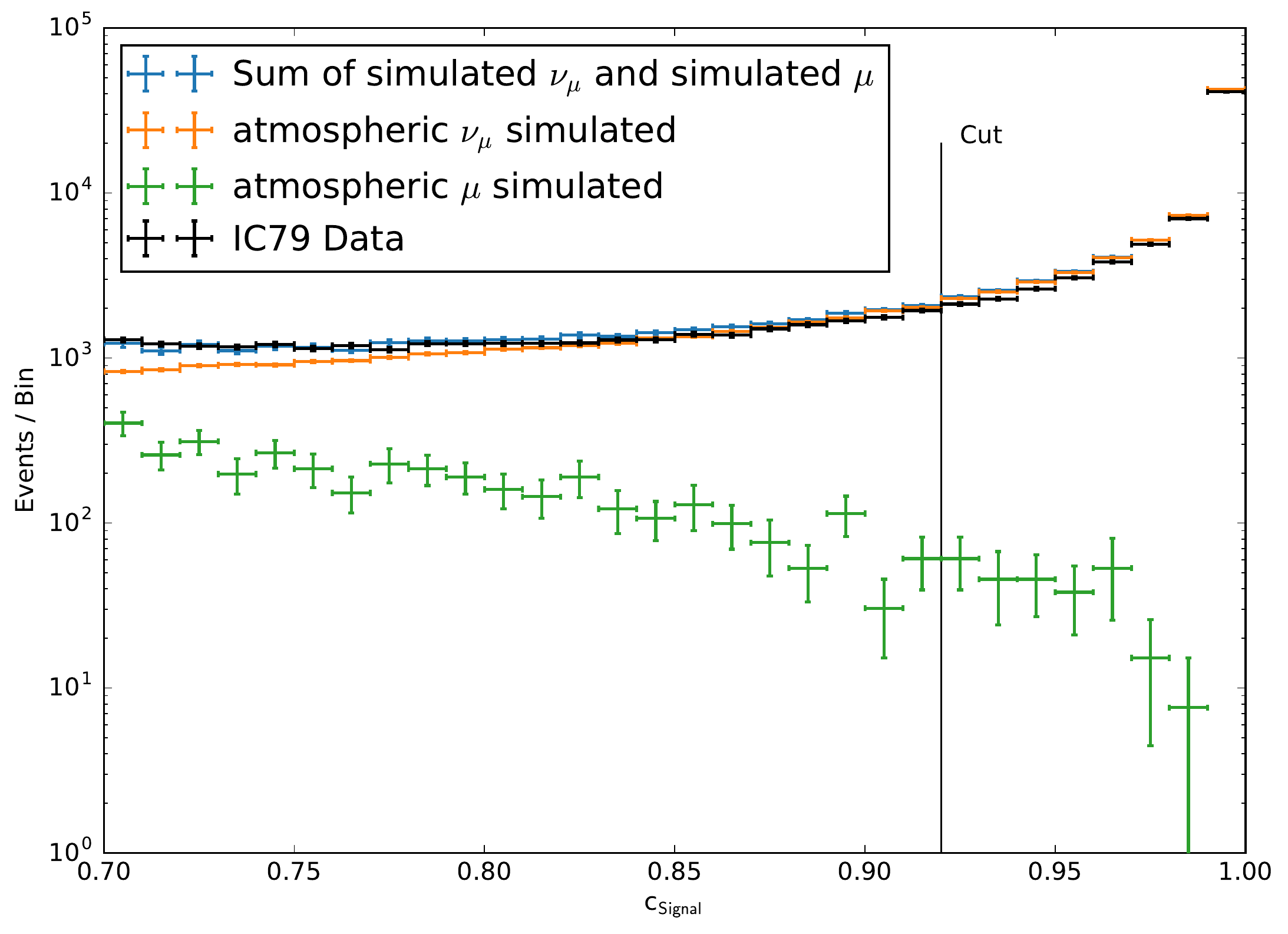}
\caption{Confidence distribution for data and simulation. Low confidence values indicate background-like events and high confidence values indicate signal-like events. A cut in the confidence $\geq$ 0.92 yields a sample with a purity of (99.5 $\pm$ 0.3)\%.}
\label{fig:confidence}
\end{figure}

In the bootstrapping, a total of 200 Random Forest models were trained, each built on a randomly chosen sample with 50\% of the size of the full sample. Using this technique, each event is scored on average 100 times. By normalizing the resulting confidence distribution for each event, the approximation of the confidence distribution is improved by taking the variance of $c_{\mathsf{Signal},i}$ into account. Furthermore, it provides statistical uncertainties for the classification. Using this way to control stability and performance, the parameters of the Random Forest were set to $k=5$ and 200 trees. The forest was trained using 120,000 simulated signal events and 30,000 simulated background events. The resulting confidence distributions for simulated events and experimental data show good compatibility and confirm a stable separation (see Fig.~\ref{fig:confidence}). No signs of overtraining were observed in the cross validation. 

%

The cut on $c_{\mathsf{Signal}}$ is a trade-off between background rejection and signal efficiency. Due to the steeply falling spectrum of atmospheric neutrinos and the expected contribution of astrophysical neutrinos, the cut was selected to yield a sufficient number of events in the highest energy bins. For this analysis, a cut at $c_{\mathsf{Signal}}\geq0.92$ was chosen (see Fig.~\ref{fig:confidence}).

This cut yields a total of 66,885 neutrino candidates in 319.6 days of detector livetime ($2.26\cdot 10^{-3}$ neutrino candidates per second). The number of background events surviving to the final level of the analysis was estimated to $330 \pm 200$ ($(1.10\pm0.73)\cdot 10^{-5}$ background events per second), which corresponds to an estimated purity of (99.5 $\pm$ 0.3)\%. In total, 21 neutrino candidates with a reconstructed muon energy $E_{\mu,\mathsf{reco}}\geq 10\,\unit{TeV}$ were observed. 

\begin{figure*}
        \centering
        \begin{subfigure}[b]{0.9\textwidth}
                \includegraphics[width=\textwidth]{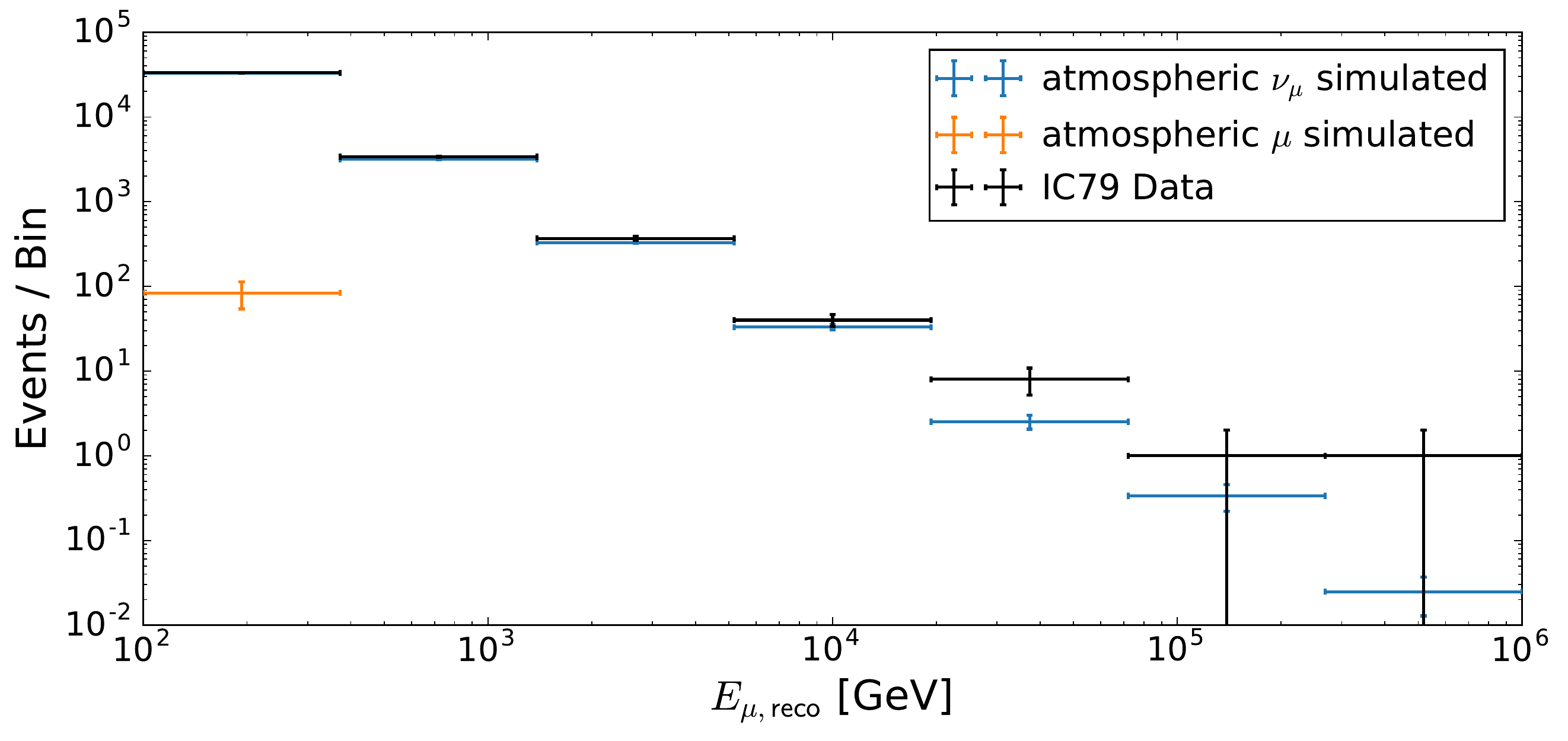}
        \end{subfigure}%

        \begin{subfigure}[b]{0.45\textwidth}
                \includegraphics[width=\textwidth]{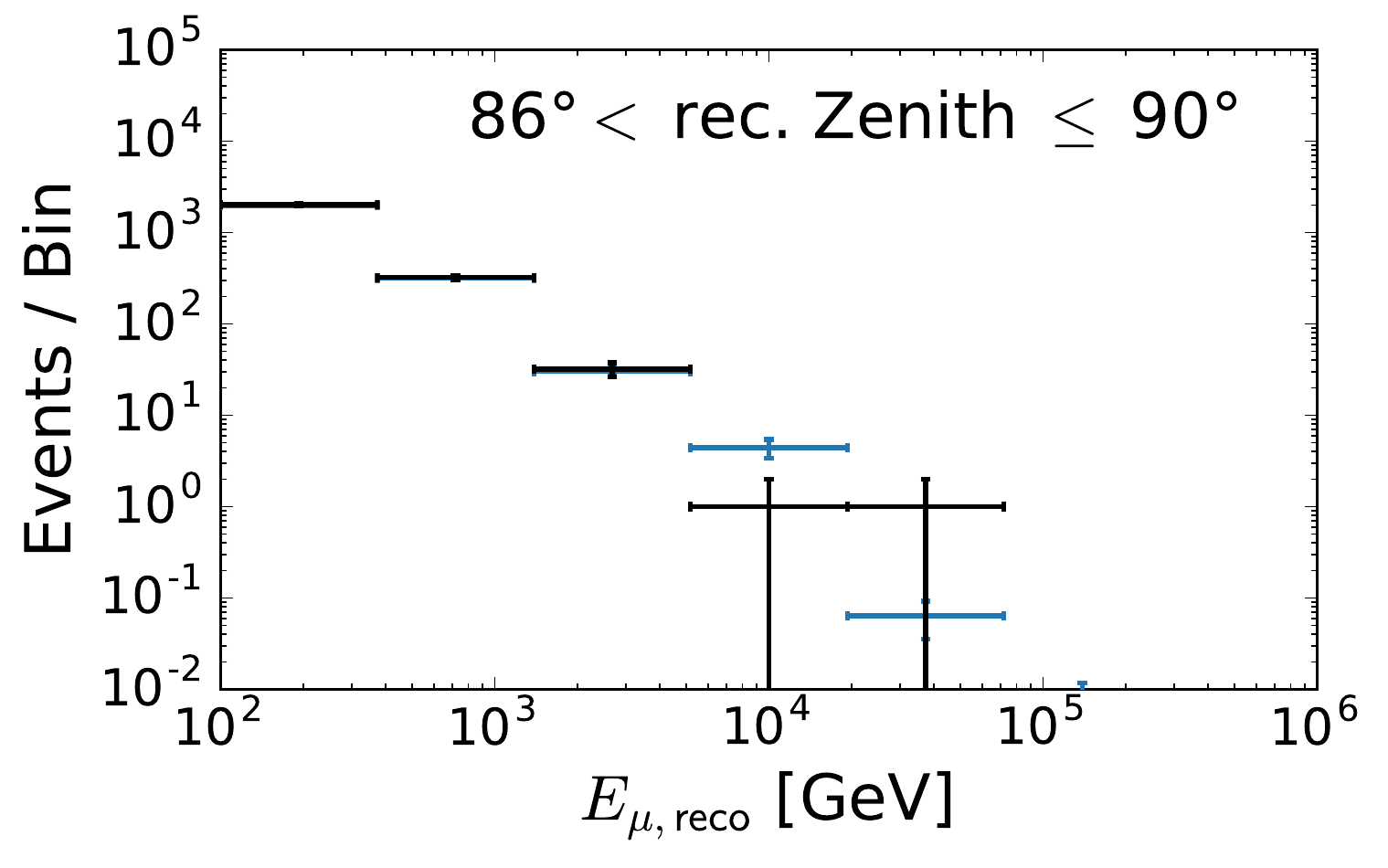}
        \end{subfigure}
        
        \begin{subfigure}[b]{0.45\textwidth}
                \includegraphics[width=\textwidth]{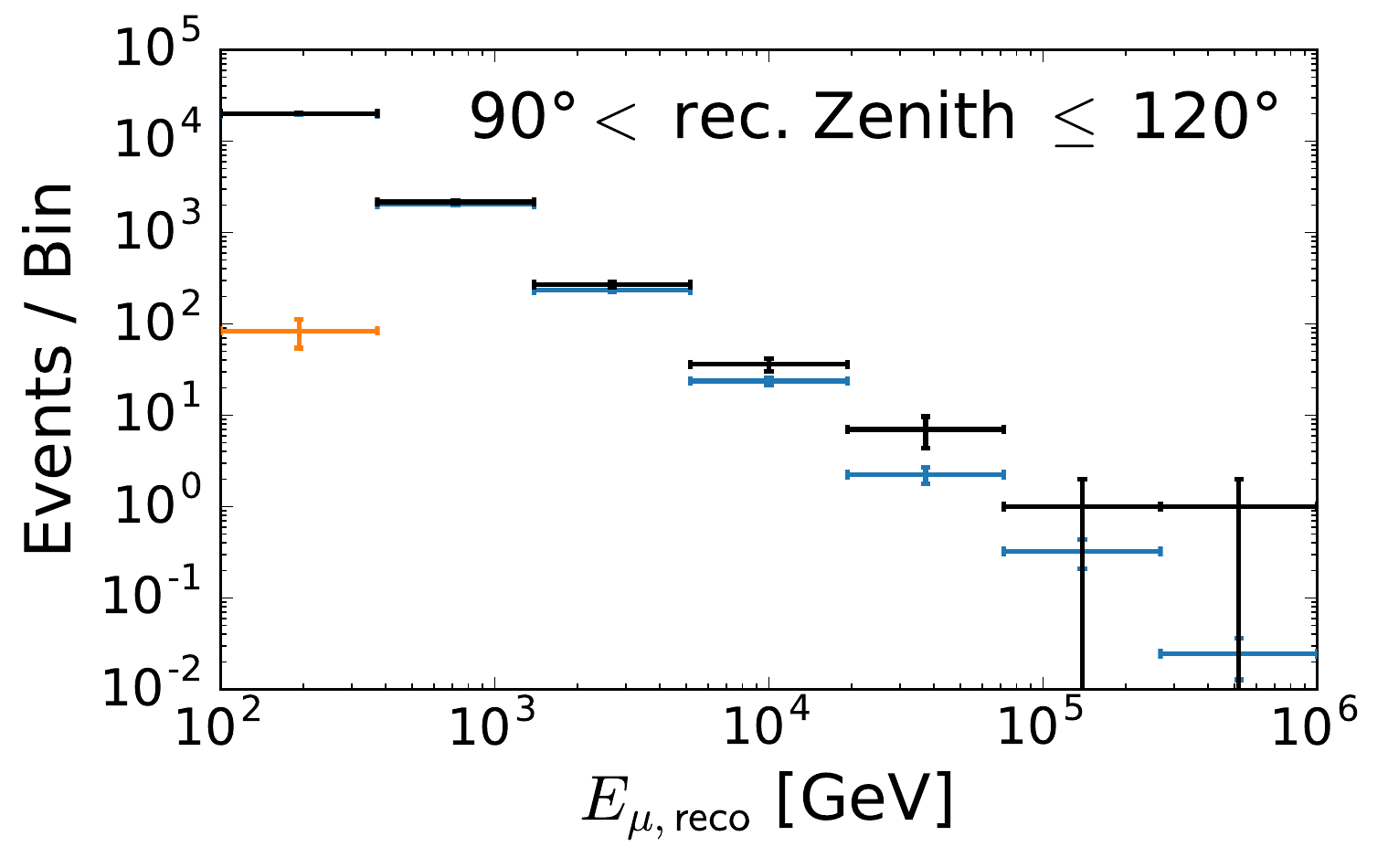}
        \end{subfigure}
        ~ 
        \begin{subfigure}[b]{0.45\textwidth}
                \includegraphics[width=\textwidth]{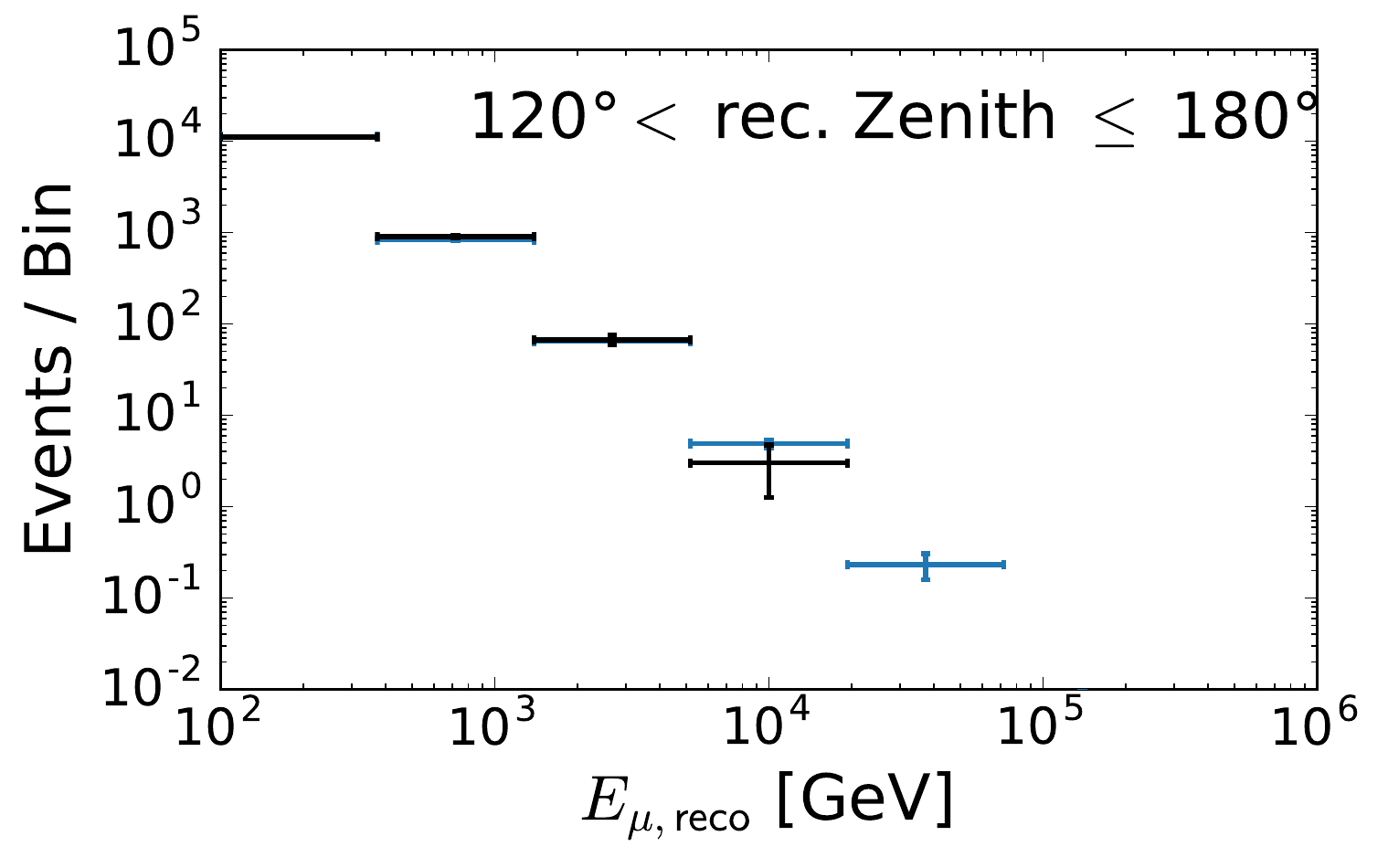}
        \end{subfigure}
        \caption{Distribution of the reconstructed muon energy. The plot in the top shows the full zenith range, while the three smaller plots show the zenith ranges indicated.}\label{fig:muex}
\end{figure*}

A good understanding of the background is mandatory to ensure that the remaining background lies in a region clearly dominated by signal events. To this end, the distributions of the reconstructed muon energy~\cite{energyreco} were investigated in different zenith regions (see Fig.~\ref{fig:muex}).
In the region from $\theta = 120\dg$ to $\theta = 180\dg$, no background and a good compatibility between data and simulation is observed. Due to earth absorption~\cite{Volkova_and_Zatsepin,Albuquerque:2001jh}, no high energy events ($E_{\mu,\mathsf{reco}}\gtrsim13\,\unit{TeV}$) are observed in experimental data in this region. Closer to the horizon ($\theta = 90\dg$to $\theta = 120\dg$), however, a few atmospheric muons are expected at the lowest energies, but this backround is more than three orders of magnitude smaller, compared to the expected number of atmospheric neutrinos and therefore negligible. Between $316\,\unit{TeV}$ and $1\,\unit{PeV}$, an excess over the atmospheric-only prediction is observed in reconstructed muon energies. 

No simulated background events are found for zenith angles $86\dg < \theta \leq 90\dg$. The livetime of experimental data, however, is about a factor of 50 larger than the livetime of the simulated background. From the results obtained on simulation, one therefore conludes that less than 54 atmospheric muon events are expected to enter the sample from this region.


\section{Unfolding}
\label{sec:Unfolding}

As the energy of the incoming neutrino cannot be accessed directly, it needs to be inferred from energy losses of the neutrino induced-muon within the detector. In the energy region of this analysis, most muon tracks are only partially contained in the detector. Furthermore, the conversion of a neutrino of energy $E_{\nu}$ into a muon of energy $E_{\mu}$ is a stochastic process. Thus, the challenge is to compute the muon neutrino energy spectrum from the reconstructed energy of the muons.

This type of problem is generally referred to as an inverse problem. It is described by the Fredholm integral equation of the first kind:
\begin{equation}
g(y)=\int_c^dA(y,x)f(x) \, \mathrm{d} x + b(y),
\label{Eq:Unfolding:fredholm}
\end{equation}
where $g(y)$ is the distribution of a measured variable $y$ and $f(x)$ is the distribution of the sought-after variable $x$. $A(y,x)$ is generally referred to as the response function and gives the probability to measure a certain $y$ given a specific $x$. The response function includes the physics of neutrino interactions, as well as the propagation of muons through the ice and all smearing effects introduced by the detector. The term $b(y)$ is the distribution of $y$ for any observed background. Due to the high purity (see Sec.~\ref{sec:EventSelection}), $b(y)$ is negligible in this analysis.

Several algorithms are available to obtain a solution to Eq.~\eqref{Eq:Unfolding:fredholm}. In the analysis presented here, the software \truee~\cite{Milke:2012ve}, which is based on the RUN algorithm~\cite{RunManual}, was used to extract the spectrum.

{\truee} allows for the use of up to three input variables and generates a binned distribution $\vec{g}$ from them. Cubic B-splines are used for the discretization of $f(x)$ and accordingly $\vec{f}$ contains the spline coefficients. The response function is transformed into a matrix $\vec{A}$ accordingly and needs to be determined from simulated events. Within {\truee}, this leads to an equation of the form:
\begin{equation}
\vec{g} = \vec{A} \vec{f}.
\label{Eq:Unfolding:Matrixeq}
\end{equation}

For most practical applications, $\vec{A}$ is ill-conditioned due to the complex mapping between $x$ and $y$. Therefore, regularization is required to avoid unstable solutions.

In this analysis, three variables were used as input to the unfolding. This is a trade-off between gaining information by using more input variables and the required number of simulated events, which grows drastically with the number of input variables. In this analysis, the reconstructed muon energy $E_{\text{reco}}$, the reconstructed track length $L_{\text{reco}}$, and the number of detected unscattered photons $N_{\text{ph}}$ were used as input to the algorithm. The three variables were mainly chosen due to their good correlation with the neutrino energy. Furthermore, the combination of these specific variables in an unfolding exhibits a positive synergy effect, which was also observed in previous unfoldings~\cite{IC59}. 

The energy proxy $E_{\text{reco}}$ is obtained by fitting the expected number of photons via an analytic template. This template scales with the energy of the muon~\cite{energyreco}. A different approach, which discards energy losses from track segments with the highest energy loss rates (generally referred to as \emph{truncated mean} $E_{\text{trunc}}$), reconstructs the muon energy more accurately~\cite{truncatedmean}. $E_{\text{reco}}$, however, was found to yield a better overall performance of the unfolding, especially when combined with $N_{\text{ph}}$ and $L_{\text{reco}}$.

\begin{figure}
 \includegraphics[width=1.\columnwidth]{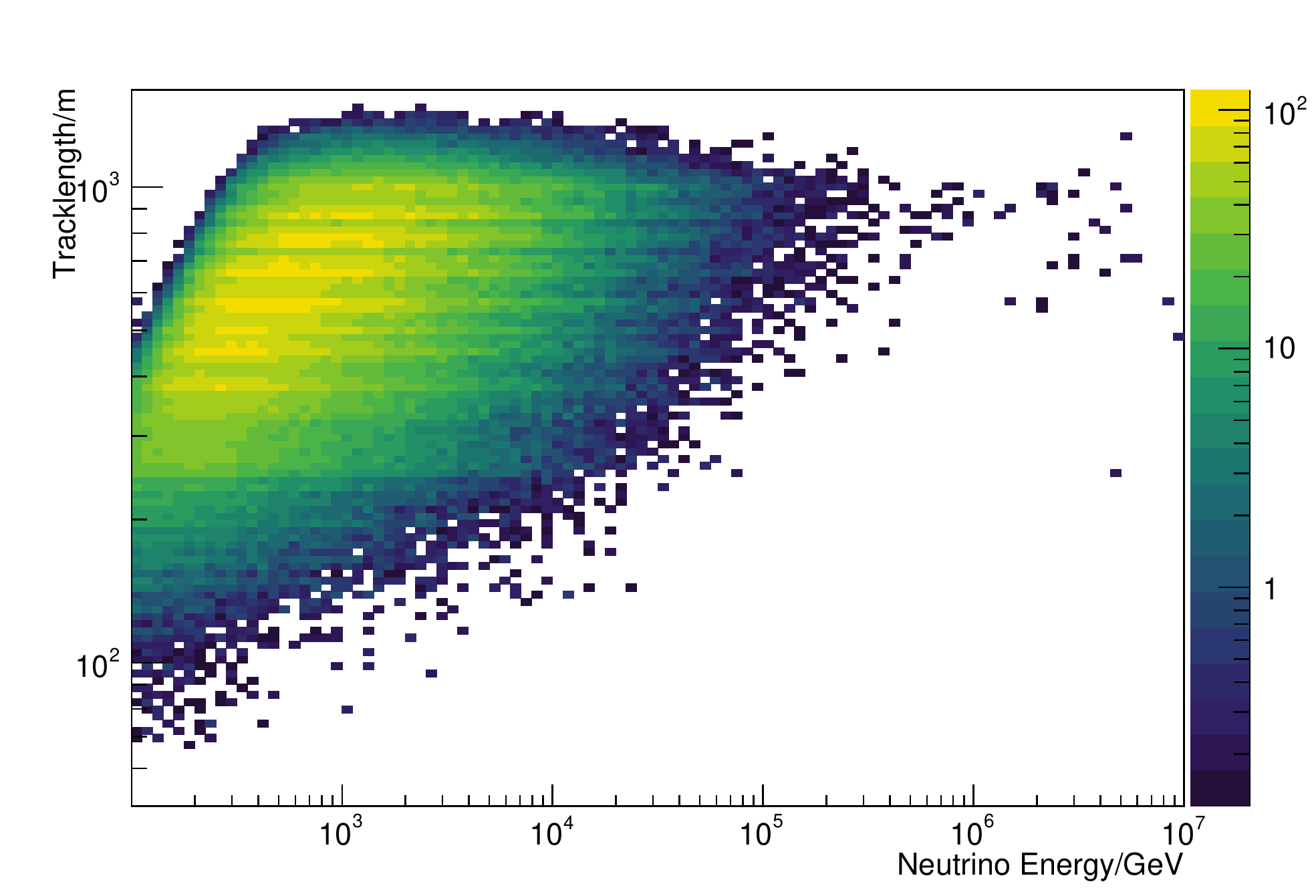}
 \caption{
  Correlation between the unfolding observable reconstructed track length and the true energy of the neutrino obtained from simulated events. The horizontal structures in this plot stem from the string spacing. 
  }
\label{fig:correlation_LDirC}
\end{figure}

\begin{figure}
\includegraphics[width=1.\columnwidth]{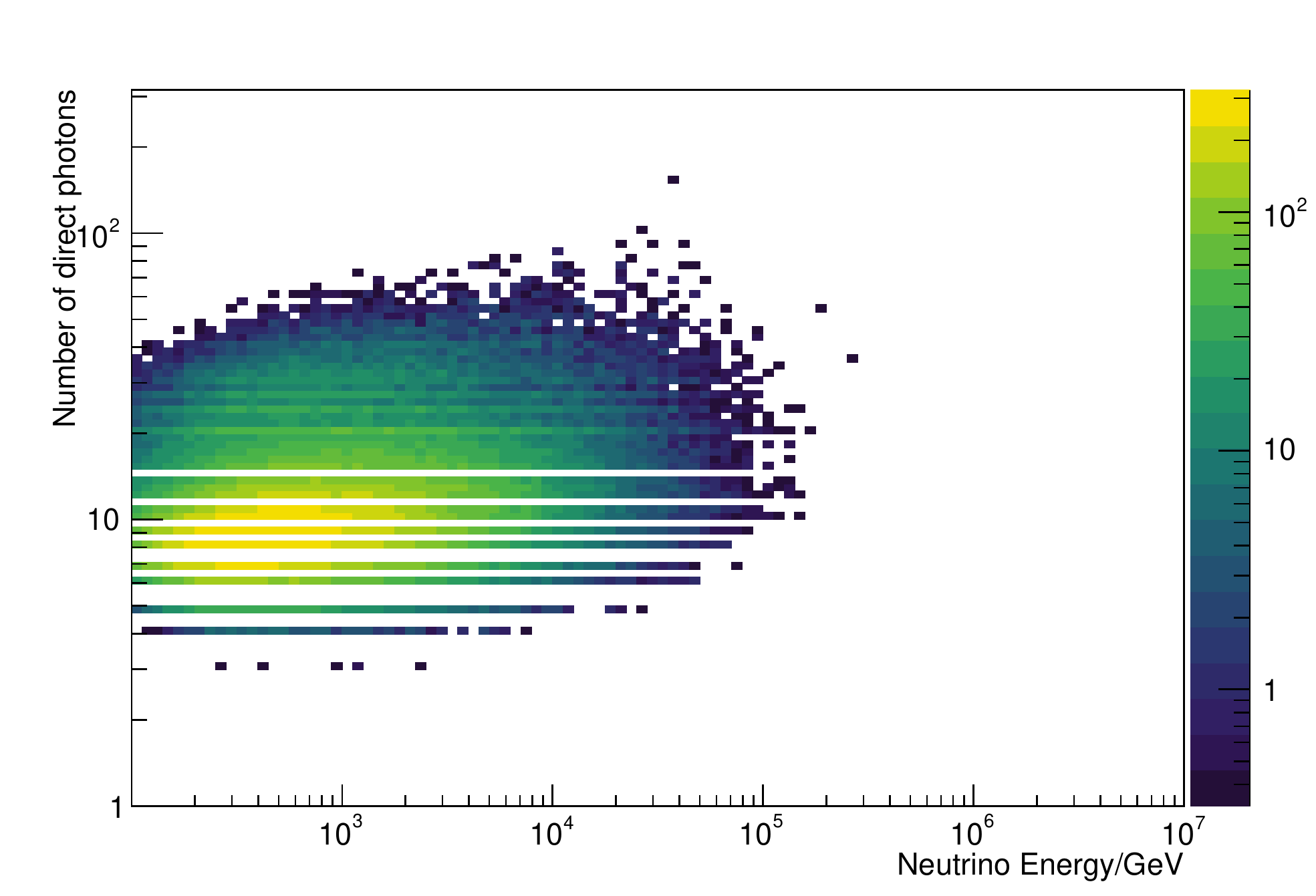}
\caption{Correlation between the unfolding observable number of direct photoelectrons measured and the true energy of the neutrino obtained from simulated events.}
\label{fig:correlation_NDirC}
\end{figure}

\begin{figure}
\includegraphics[width=1.\columnwidth]{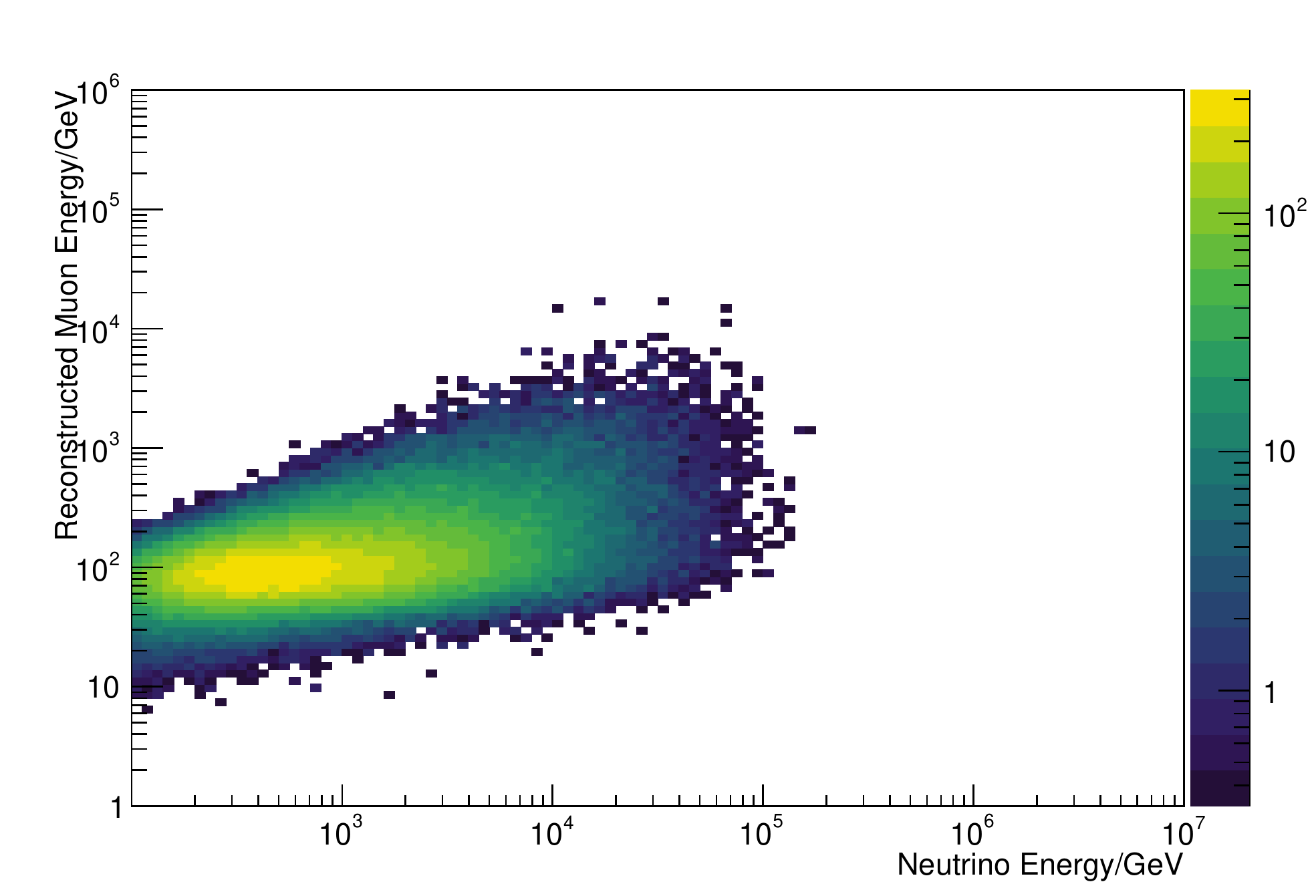}
\caption{Correlation between the unfolding observable reconstructed muon energy at the center of the detector and the true energy of the neutrino obtained on simulated events.}
\label{fig:correlation_MuEX}
\end{figure}

The correlation of the individual variables with the energy of the incoming neutrino is depicted in Figs.~\ref{fig:correlation_LDirC}~to~\ref{fig:correlation_MuEX}. All three input variables are strongly correlated with energy and were also used in \cite{IC59}.  The horizontal bands in Fig.~\ref{fig:correlation_LDirC} arise from the fact that certain track lengths are preferred in the reconstruction, which is due to the integer number of strings and the regularity of the array. A detailed description of {\truee} and its implemented validation methods can be found in~\cite{Milke:2012ve}, while its application in a spectral measurement is described in~\cite{IC59}.
\begin{figure*}[htb]
\centering
\includegraphics[width=0.90\textwidth]{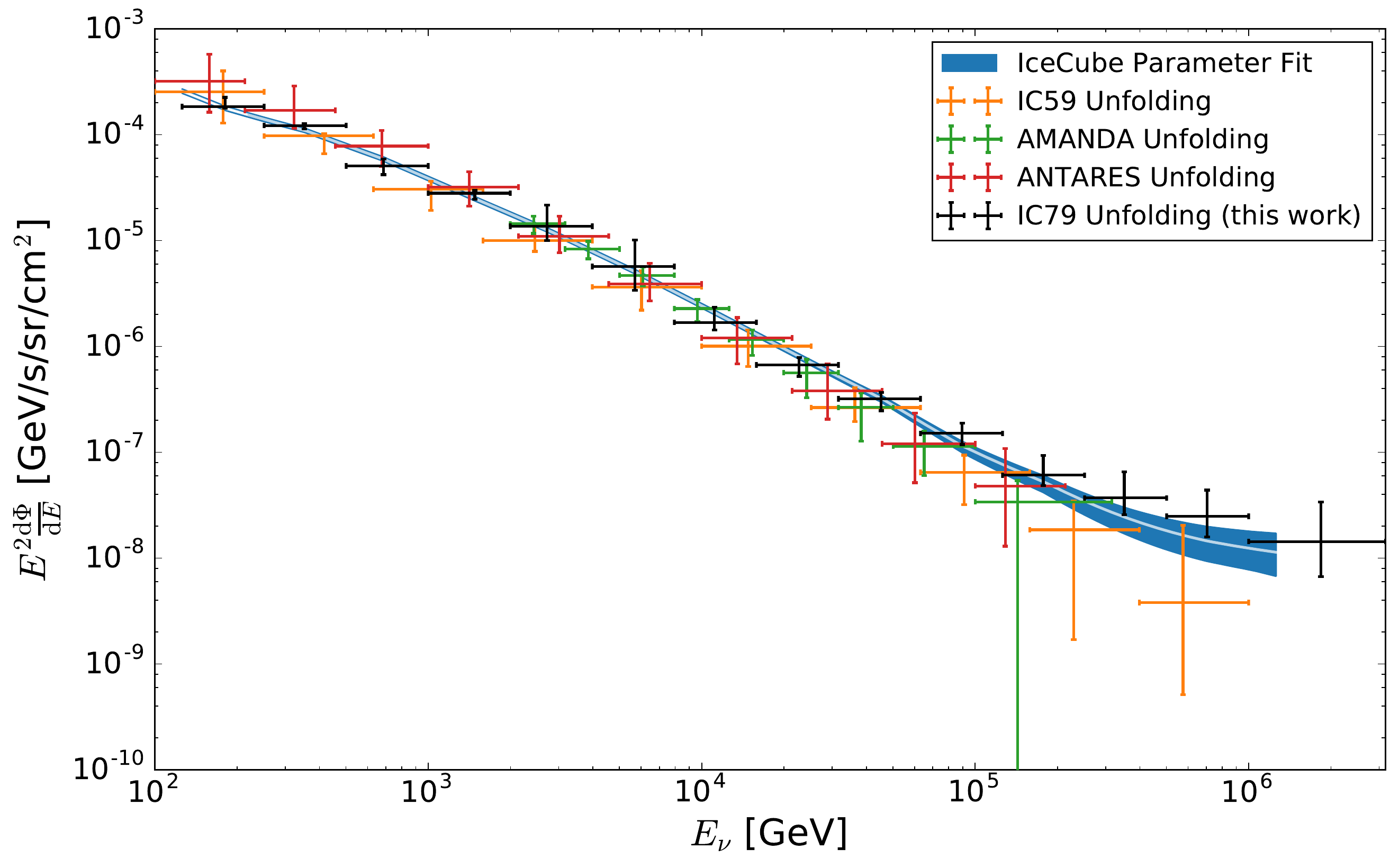}
\caption{The obtained $\nu_\mu$ spectrum of this analysis compared to the unfolding analyses of AMANDA \cite{amanda}, ANTARES \cite{antares} and IceCube-59 \cite{IC59}. The unfoldings can have slightly different zenith dependent sensitivities. In addition to the unfolding results the best fit and its uncertainties from an IceCube parameter fit~\cite{ic79weaver}, evaluated for the zenith dependent sensitivity of this work, are shown.}
\label{fig:experimentalcomparison}
\end{figure*}
\begin{figure*}[htb]
\centering
\includegraphics[width=0.90\textwidth]{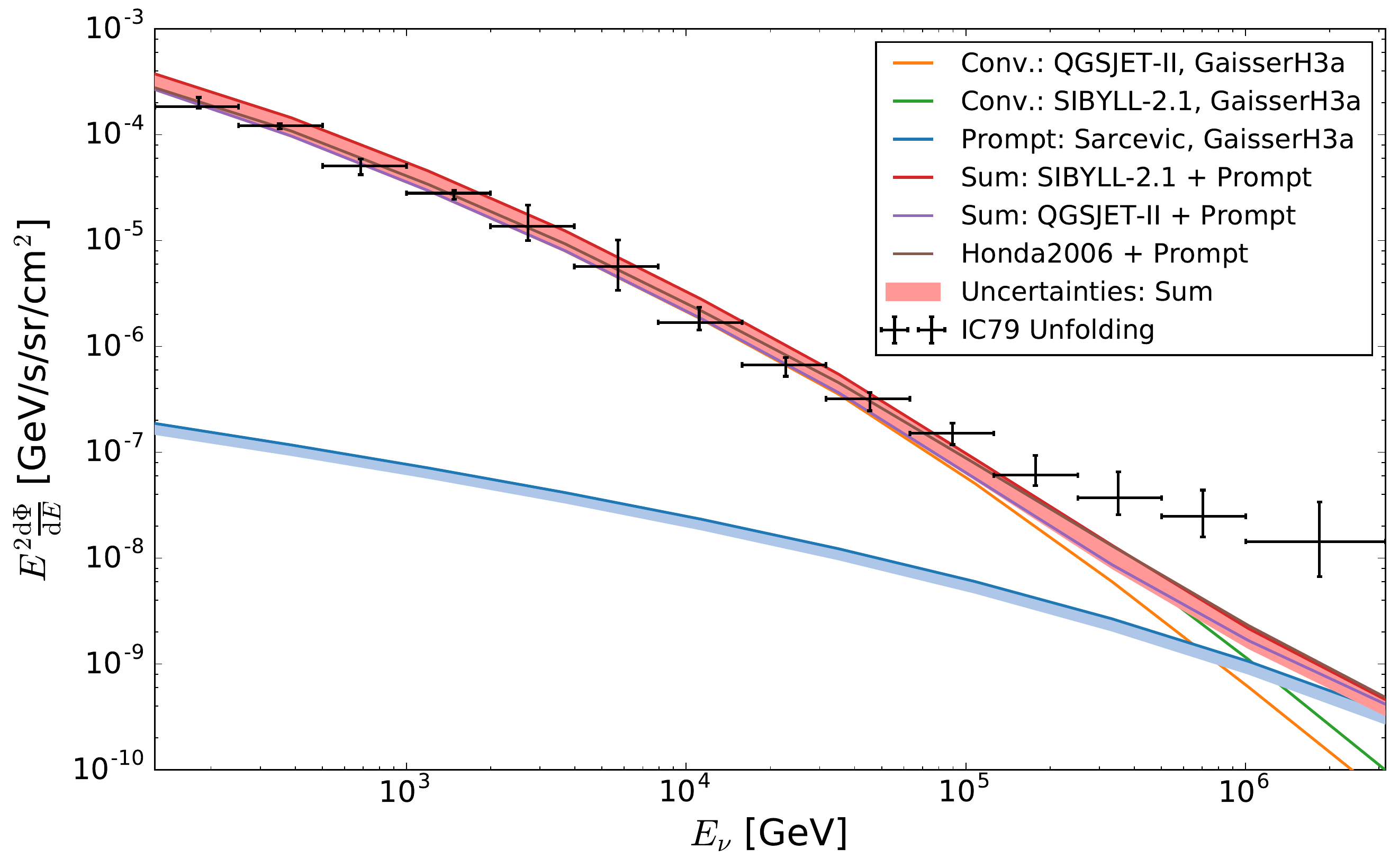}
\caption{Unfolded $\nu_\mu$ energy spectrum compared to theoretical calculations. The conventional models Sybill-2.1 and QGSJET-II are used as upper and lower bounds for possible models~\cite{Anatoli}. Most of the common models such as \cite{Honda} lie in between those two models over the whole energy range. For the prompt component, the flux from \cite{Enberg} is used. The blue shaded area represents the theoretical uncertainty on the prompt flux as reported in~\cite{Enberg}. The pink shaded area depicts the sum of uncertainties arising from the conventional and prompt components, respectively. All predictions are calculated for the primary spectrum proposed in \cite{Gaisser2012801} and for the zenith dependent sensitivity shown in Tab.~\ref{tab:acceptance}.}
\label{fig:theoreticalcomparison}
\end{figure*}

In spectral measurements with IceCube, systematic uncertainties are dominated by two sources: The first source is associated with the amount of light detected in an event, which is affected by the light detection efficiency of the optical modules and by the muon interaction cross-sections (ionization, pair-production and photonuclear interaction). These two uncertainties cannot be disentangled on experimental data and are therefore combined in a single value associated with the efficiency of the DOMs. The common calibration error on the photomultiplier efficiency is 7.7\%~\cite{DOMCalibration}, whereas the theoretical uncertainty on the muon cross-sections was estimated to 4\% in~\cite{Kokoulin1999}. The second source of systematic uncertainties is associated with the scattering and absorption of photons in the glacial ice at the South Pole~\cite{propofice,ice2013}. The results of the natural formation of the detection medium are inhomogeneities that are accounted for by systematically changing the scattering and absorption lengths in Monte Carlo simulations. 

As the unfolding result is based on simulated events which are used as input to TRUEE, the specific choices (ice-model, DOM efficiency) used in the generation of these events affect the outcome of the spectral measurement. To estimate the impact of changes with respect to the so-called baseline simulation, systematic uncertainties were derived using a bootstrapping procedure. Within this bootstrapping, a number of simulated events corresponding to the number of neutrino candidates on experimental data was drawn at random from a systematic data set. 

In total, five systematic sets were available for the analysis at hand. In each of these sets, one property has been varied with respect to the baseline simulation. In one set, the efficiency for all DOMs was increased by 10\%, while in a second, the efficiency for all DOMs was decreased by 10\%. To account for uncertainties in the description of the ice, three sets of simulated events with different ice models were generated. The first set was produced with a scattering length increased by 10\% with respect to the baseline simulation, whereas the second one was produced with an absorption length also increased by 10\% with respect to the baseline simulation. In a third set, the effects of scattering and absorption were combined by simultaneously decreasing scattering and absorption length by 7.1\%. 

Based on these simulated events the energy spectrum was obtained, while using the baseline simulation for the extraction of the response matrix. In a next step, the discrepancy between the unfolding result and the true distribution was computed. This procedure was repeated 1,000 times on any of the five systematic data sets and yielded the contribution of the individual sources of uncertainty. The total systematic uncertainty was computed as the sum of squares of the individual contributions. Compared to the statistical uncertainty, the systematic uncertainty is found to be large, except for the last two bins, where both uncertainties are of approximately the same size. 

The unfolded flux of atmospheric muon neutrinos, including statistical and systematic uncertainties, is summarized in Tab.~\ref{tab:Flux}. A comparison to previous measurements~\cite{amanda,IC59,ic79weaver,antares} is depicted in Fig.~\ref{fig:experimentalcomparison}. No significant difference to any of the depicted measurements is observed over the entire energy range.

A comparison to theoretical predictions for the atmospheric neutrino flux is shown in Fig.~\ref{fig:theoreticalcomparison}. A clear excess over theoretical predictions is observed for the last four bins which cover an energy range from $E_\nu = 126\,\unit{TeV}$ to $E_\nu=3.2\,\unit{PeV}$. The size of the excess was computed with respect to the Honda 2006 model~\cite{Honda2006}, as well as with respect to fluxes of atmospheric neutrinos predicted using the cosmic ray interaction models SIBYLL-2.1~\cite{SIBYLL} and QGSJET-II~\cite{QGSJET} and is found to vary depending on energy and the underlying theoretical model. 

The largest deviation of $2.8\,\sigma$ is observed in the 11-th bin ($E_{\text{Center}}=177.8\,\unit{TeV}$) with respect to an atmospheric neutrino flux computed using QGSJET-II. For atmospheric neutrino fluxes computed using SIBYLL-2.1 and the Honda 2006 model, the largest excess of $2.4\,\sigma$ is observed for the 13-th bin ($E_{\text{Center}}=707.9\,\unit{TeV}$). For all three flux models discussed above, the prompt component of the atmospheric flux was modeled according to the ERS model~\cite{Enberg}. Uncertainties on the theoretical predictions -- indicated by the shaded bands in Fig.~\ref{fig:theoreticalcomparison} -- were not taken into account for the calculation of the excess. The size of the excess is found to decrease for the last bin, due to larger statistical and systematic uncertainties. Table~\ref{tab:significance} summarizes the significance of the excess for the four highest energy bins with respect to selected model calculations. A slight increase of the significances is found when results of recent perturbative QCD calculations~\cite{berss,Garzelli2015,Gauld2016} are used to model the prompt component. 

Due to the relatively small number of events observed in the four highest energy bins, the observed correlation between those bins is rather large, especially in the case of neighbouring bins. This prohibits an accurate estimation of the spectral index of the diffuse flux of high energy astrophysical neutrinos, as well as its normalization. Furthermore, in such an estimate, two observables would be estimated on the basis of only four data points. 

\begin{table*}[htb]
\centering
\caption{Data points and uncertainties of the unfolding results.}
\label{tab:Flux}
\begin{tabular}{cccccccc}
\toprule
Energy range & Center Energy & Flux & \multicolumn{2}{c}{Total Uncertainty} & Stat. Uncertainty & Cov. Matrix (Diag. El.)\\
$[\log_{10}(\textrm{E}/\textrm{GeV})]$ &
$[\log_{10}(\textrm{E}/\textrm{GeV})]$ &
$[\textrm{E}^2\frac{\text{GeV}}{\text{sr}\,\text{s}\,\text{cm}^2}]$ &
 &  & \\
\midrule
$2.10 - 2.40$ & $2.26$ & $1.84\cdot 10^{-4}$ & $ +22\%$  &$-3\% $   & $\pm3\%$ & $2.78\cdot10^{-20}$  \\
$2.40 - 2.70$ & $2.55$ & $1.22\cdot 10^{-4}$ & $  +5\%$  &$-7\% $   & $\pm2\%$ & $3.76\cdot10^{-20}$ \\
$2.70 - 3.00$ & $2.84$ & $5.07\cdot 10^{-5}$ & $ +16\%$  &$-17\%$   & $\pm4\%$ & $1.80\cdot10^{-23}$ \\
$3.00 - 3.30$ & $3.17$ & $2.80\cdot 10^{-5}$ & $  +6\%$  &$-12\%$   & $\pm4\%$ & $2.62\cdot10^{-25}$ \\
$3.30 - 3.60$ & $3.43$ & $1.37\cdot 10^{-5}$ & $ +58\%$  &$-27\%$   & $\pm4\%$ & $5.72\cdot10^{-27}$ \\
$3.60 - 3.90$ & $3.76$ & $5.69\cdot 10^{-6}$ & $ +78\%$  &$-40\%$   & $\pm5\%$ & $7.38\cdot10^{-29}$ \\
$3.90 - 4.20$ & $4.05$ & $1.68\cdot 10^{-6}$ & $ +39\%$  &$-15\%$   & $\pm7\%$ & $8.73\cdot10^{-31}$ \\
$4.20 - 4.50$ & $4.36$ & $6.69\cdot 10^{-7}$ & $ +17\%$  &$-22\%$   & $\pm10\%$ & $1.62\cdot10^{-32}$ \\
$4.50 - 4.80$ & $4.66$ & $3.20\cdot 10^{-7}$ & $ +15\%$  &$-23\%$   & $\pm10\%$ & $2.35\cdot10^{-34}$\\
$4.80 - 5.10$ & $4.95$ & $1.51\cdot 10^{-7}$ & $ +25\%$  &$-22\%$   & $\pm13\%$ & $6.11\cdot10^{-36}$\\
$5.10 - 5.40$ & $5.25$ & $6.08\cdot 10^{-8}$ & $ +53\%$  &$-20\%$   & $\pm18\%$ & $1.20\cdot10^{-37}$\\
$5.40 - 5.70$ & $5.55$ & $3.71\cdot 10^{-8}$ & $ +76\%$  &$-31\%$   & $\pm25\%$ & $5.43\cdot10^{-39}$\\
$5.70 - 6.00$ & $5.85$ & $2.48\cdot 10^{-8}$ & $ +77\%$  &$-36\%$   & $\pm35\%$ & $3.00\cdot10^{-40}$\\
$6.00 - 6.50$ & $6.26$ & $1.44\cdot 10^{-8}$ & $+136\%$  &$-53\%$   & $\pm53\%$ & $5.31\cdot10^{-42}$\\ 
\bottomrule
\end{tabular}
\end{table*}
\begin{table}[htb]
\centering
\caption{Sensitivity of the detector for different energies and zenith regions. The percentages are the share of events detected from a zenith region for a selected energy bin, assuming a uniformly distributed flux in $\cos(\Theta)$. Without earth absorption and with an ideal detector, the sensitivity would be $25\%$ in each zenith bin. The selected zenith bins are: $\Theta_{\text{North}} \in \left[\cos{180^\circ},\cos{137^\circ}\right]$, $\Theta_{1} \in \left[\cos{137^\circ},\cos{117^\circ}\right]$, $\Theta_{2} \in \left[\cos{117^\circ},\cos{101^\circ}\right]$ and $\Theta_{\text{Horizon}} \in \left[\cos{101^\circ},\cos{86^\circ}\right]$. }
\label{tab:acceptance}
\begin{tabular}{ccccc}
\toprule
Energy range & \multicolumn{4}{c}{$N_{i}/N_{\text{Uniform Flux}}$} \\
$[\log_{10}(\textrm{E}/\textrm{GeV})]$ &
$\Theta_{\text{North}}$ &
$\Theta_{1}$ &
$\Theta_{2}$ &
$\Theta_{\text{Horizon}}$\\
\midrule
$2.10 - 2.40$ & $24.4\%$ & $25.6\%$ & $29.0\%$ & $21.0\%$ \\
$2.40 - 2.70$ & $20.4\%$ & $24.2\%$ & $29.1\%$ & $26.4\%$ \\
$2.70 - 3.00$ & $18.1\%$ & $22.7\%$ & $29.2\%$ & $30.0\%$ \\
$3.00 - 3.30$ & $15.9\%$ & $21.5\%$ & $29.7\%$ & $32.9\%$ \\
$3.30 - 3.60$ & $13.7\%$ & $19.4\%$ & $29.7\%$ & $37.2\%$ \\
$3.60 - 3.90$ & $11.3\%$ & $17.1\%$ & $29.4\%$ & $42.3\%$ \\
$3.90 - 4.20$ & $ 9.4\%$ & $15.4\%$ & $28.3\%$ & $46.9\%$ \\
$4.20 - 4.50$ & $ 8.1\%$ & $13.1\%$ & $27.2\%$ & $51.7\%$ \\
$4.50 - 4.80$ & $ 5.8\%$ & $11.4\%$ & $25.5\%$ & $57.3\%$ \\
$4.80 - 5.10$ & $ 5.5\%$ & $11.5\%$ & $24.3\%$ & $58.7\%$ \\
$5.10 - 5.40$ & $ 3.0\%$ & $10.5\%$ & $24.5\%$ & $62.0\%$ \\
$5.40 - 5.70$ & $ 2.3\%$ & $ 7.9\%$ & $24.7\%$ & $65.1\%$ \\
$5.70 - 6.00$ & $ 2.0\%$ & $ 8.1\%$ & $26.7\%$ & $63.2\%$ \\
$6.00 - 6.50$ & $ 0.8\%$ & $ 3.0\%$ & $25.9\%$ & $70.3\%$ \\
\bottomrule
\end{tabular}
\end{table}

\begin{table}[htb]
\centering
\caption{Significance of the observed excess in the four highest energy bins, computed with respect to selected theoretical predictions. For all three models discussed here, the prompt component was modelled according to the ERS model~\cite{Enberg}.}
\label{tab:significance}
\begin{tabular}{lccc}
\toprule
$E_{\text{Center}}$ in TeV & Honda 2006 & QGSJET-II & SIBYLL-2.1 \\
\midrule 
177.8 & 1.8\,$\sigma$ & 1.8\,$\sigma$ & 2.8\,$\sigma$ \\   
354.8 & 2.2\,$\sigma$ & 2.2\,$\sigma$ & 2.5\,$\sigma$ \\
707.9 & 2.4\,$\sigma$ & 2.4\,$\sigma$ & 2.5\,$\sigma$ \\
1819.7 & 1.7\,$\sigma$ & 1.7\,$\sigma$ & 1.8\,$\sigma$\\
\bottomrule
\end{tabular}
\end{table}


\section{Consistency Check}
\label{sec:Discussion}

A comparison of the unfolded spectrum with theoretical predictions for a purely atmospheric flux shows good compatibility up to energies of $E_{\nu} \approx 126\,\unit{TeV}$ (see Fig.~\ref{fig:theoreticalcomparison}). Due to lower maximum energies~\cite{amanda,antares}, larger uncertainties and the detector geometry~\cite{IC59}, no hints for a non-atmospheric component were observed in previous spectral measurements. Nevertheless, those measurements are in good agreement with the result of this analysis (see Fig.~\ref{fig:experimentalcomparison}). 

For energies above $126\,\unit{TeV}$, however, a flattening of the unfolded spectrum is observed. To verify that this flattening is consistent with an astrophysical contribution to the overall spectrum of muon neutrinos, the unfolding result is compared to previous measurements of the astrophysical flux with IceCube. 

\begin{figure*}[ttt]
        \begin{subfigure}[b]{0.45\textwidth}
                \includegraphics[width=\textwidth]{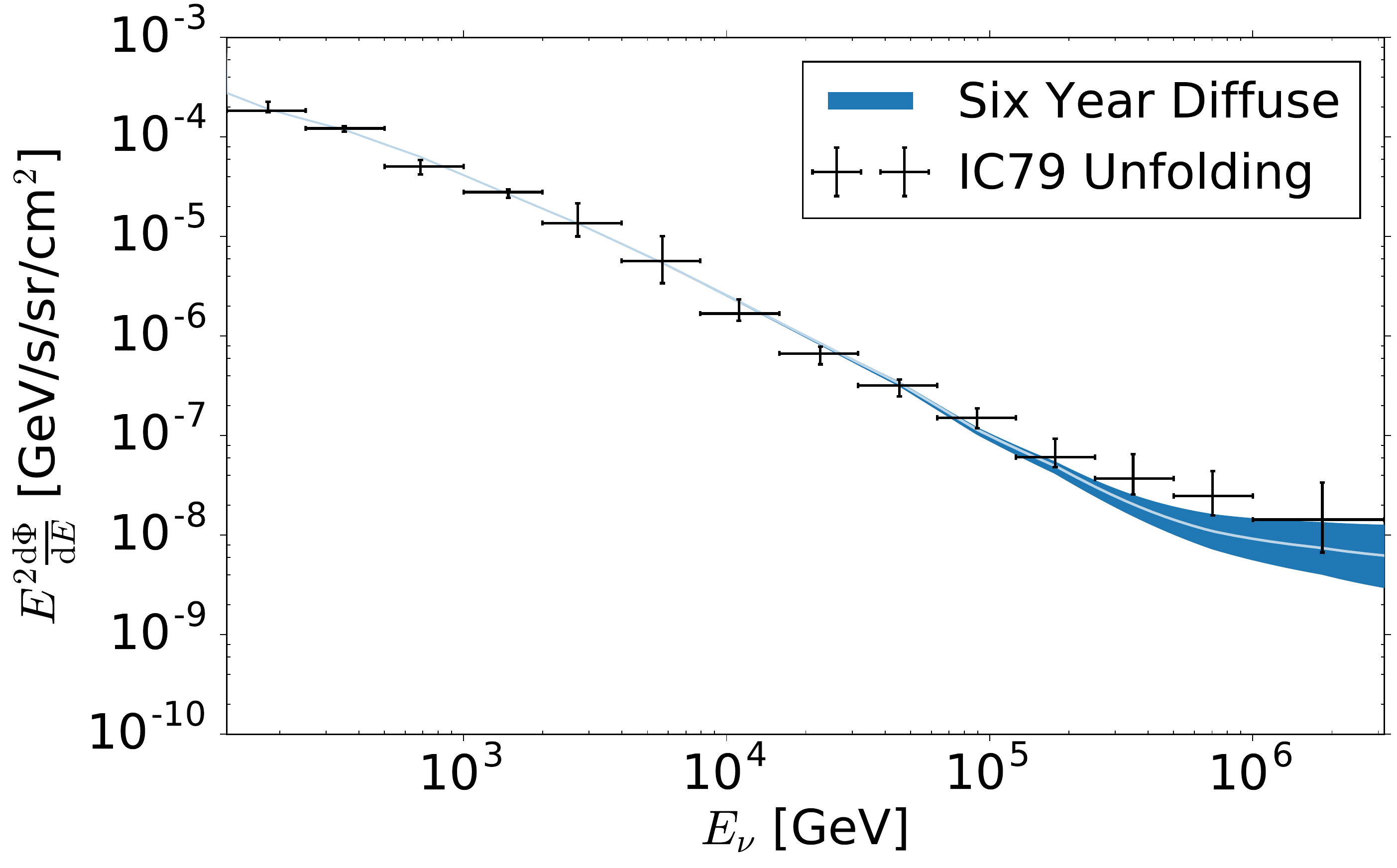}
                \caption{Comparison of the unfolded overall spectrum of muon neutrinos to a likelihood analyis of six years of IceCube data~\cite{Aachen6yrs}.}
                \label{fig:unfolding_vs_SixYearDiffuse}
        \end{subfigure}
		~
        \begin{subfigure}[b]{0.45\textwidth}
                \includegraphics[width=\textwidth]{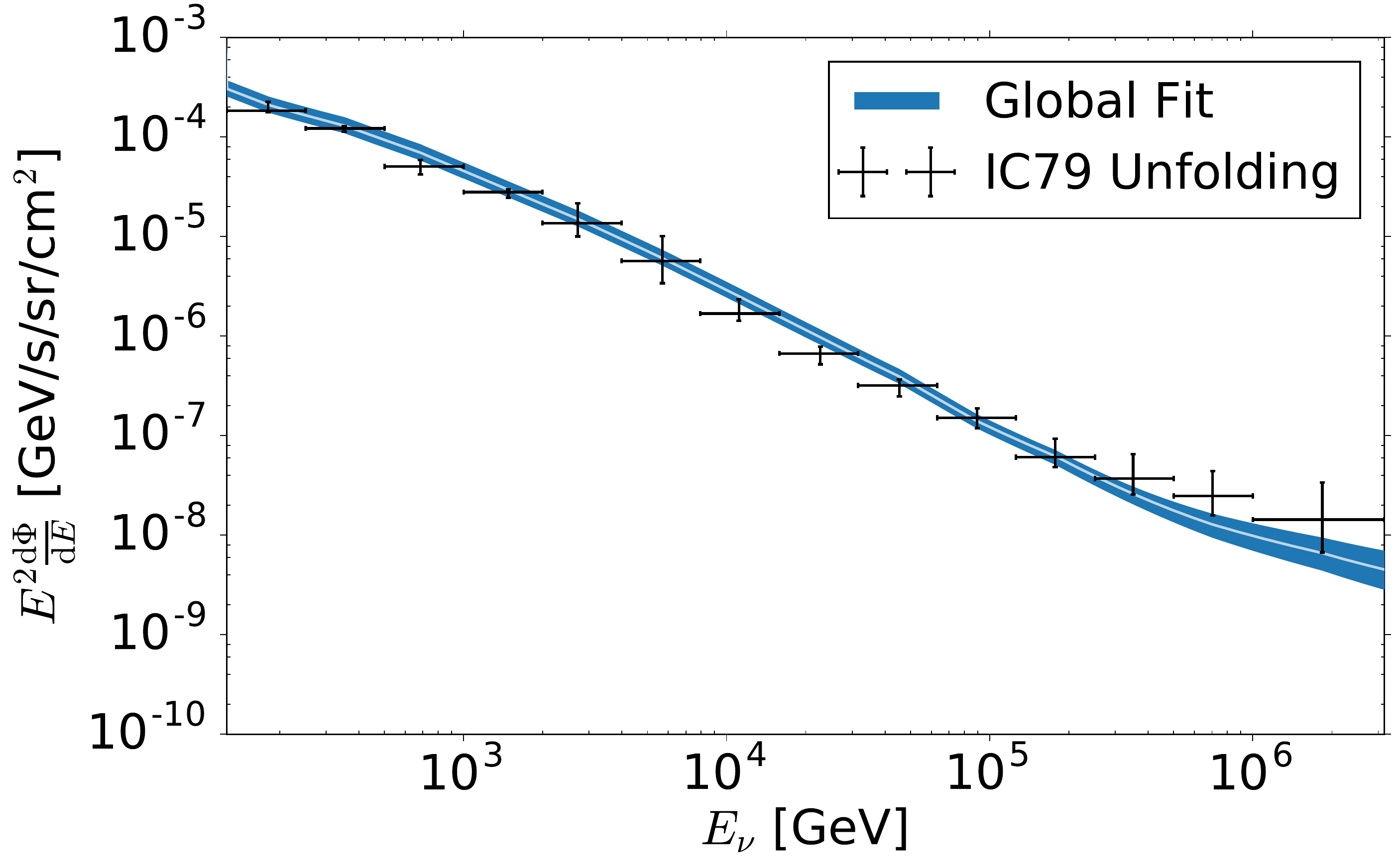}
                \caption{Comparison of the unfolded overall flux of muon neutrinos to a combined likelihood analysis of several years of IceCube data~\cite{globalfit}.}
                \label{fig:unfolding_vs_GlobalFit}
        \end{subfigure}
        \caption{Comparison of the unfolded data points to previous measurements of the astrophysical neutrino flux with IceCube. The blue shaded area represents the error band on the individual fluxes, which was derived using the 68\% C.L. on the best fit values on $\gamma_{\text{astro}}$, $\Phi_{\text{astro}}$ as well as on the contribution of conventional and prompt atmospheric neutrinos.}
\label{fig:unfolding_vs_IceCube_with_errorband}
\end{figure*}
Figure~\ref{fig:unfolding_vs_IceCube_with_errorband} depicts a comparison of the unfolded data points with measurements of the astrophysical flux by IceCube. A comparison of the unfolded data points to a likelihood analysis of muon neutrinos using six years of detector livetime~\cite{Aachen6yrs} is depicted in Fig.~\ref{fig:unfolding_vs_SixYearDiffuse}. Figure~\ref{fig:unfolding_vs_GlobalFit} compares the unfolding result to a combined likelihood analysis of several years of IceCube data~\cite{globalfit}. In each of the figures, the blue shaded area represents the uncertainty band of the respective analysis. For each measurement, the uncertainty band was computed using the 68\% confidence level errors on the best fit values for $\gamma_{\text{astro}}$, $\Phi_{\text{astro}}$ as well as on the contribution of conventional and prompt atmospheric neutrinos. Other IceCube analyses also performed measurements of the astrophysical neutrino flux~\cite{ic79weaver,aartsen2015atmospheric,icrc}. We do not, however, explicitly compare those to the unfolding result, as the obtained indices and normalizations are bracketed by the results from~\cite{Aachen6yrs} ($\gamma = 2.13 \pm 0.13$) and~\cite{globalfit} ($\gamma = 2.50\pm0.09$).

From Fig.~\ref{fig:unfolding_vs_IceCube_with_errorband}, one finds that the result of the unfolding agrees well with previous measurements of the astrophysical neutrino flux with IceCube. There is a slight disagreement between the unfolding result and the maximal flux obtained in~\cite{globalfit}. This disagreement, however, is only observed for data points below $\approx 60$\,TeV and arises from the fact that a rather large normalization of the conventional atmospheric $\nu_{\mu}$ flux ($1.1\pm0.21$ times Honda 2006) was obtained in~\cite{globalfit}. 

We therefore conclude that the flattening of the muon neutrino energy spectrum at energies above $\approx 60$\,TeV is consistent with an astrophysical flux of neutrinos. Note, however, that due to the rather large uncertainties, the unfolding cannot discriminate between the results obtained in~\cite{globalfit} and~\cite{Aachen6yrs}.


\section{Conclusion}
\label{sec:Conclusion}

In this paper, an unfolding of the \numu energy spectrum obtained with IceCube in the 79-string configuration of the detector has been presented. The unfolded spectrum covers an energy range from $125\,\unit{GeV}$ to $3.2\,\unit{PeV}$, thus extending IceCube's reach in spectral measurements by more than a factor of 3, compared to previous analyses~\cite{IC59}. 

The unfolding is based on a dataset with a high purity of (99.5 $\pm$ 0.3)\% at an event rate of $2.26\cdot 10^{-3}$ neutrino candidates per second. This is an improvement of the event rate by a factor of 2.43 compared to the previous analysis~\cite{IC59} ($0.93\cdot 10^{-3}$ neutrino candidates per second) at an equally high purity of the final sample of neutrino candidates. The improvement is the result of two steps that were altered compared to the previous analysis~\cite{IC59}. 

Firstly, the quality criteria for tracks were chosen to be less exclusive, which results in a larger number of neutrino candidates available for a further selection with machine learning techniques. This mainly results from the more symmetric shape of the IC-79 detector, when compared to IC-59. Secondly, using a bootstrapping technique, the probability density function of contaminating muon events was estimated more accurately. Therefore, the final cut on the confidence distribution (Fig.~\ref{fig:confidence}) was chosen more precisely, which results in a larger number of neutrino candidates at the final analysis level.

The distribution of the reconstructed muon energy in the final sample (Fig.\ref{fig:muex}) shows an excess of experimental data over the atmospheric prediction for reconstructed muon energies of $E_\nu > 40\,\unit{TeV}$. 

The excess at high energies was confirmed in the subsequent unfolding. The unfolded spectrum shows good compatibility with the atmospheric predictions up to neutrino energies of $\sim126\,\unit{TeV}$ (see Fig.~\ref{fig:theoreticalcomparison}). For higher energies the spectrum exceeds an atmospheric only prediction. This excess is compatible with recent measurements of an astrophysical neutrino flux (see Fig.~\ref{fig:experimentalcomparison} and Sec.~\ref{sec:Discussion}).

This analysis presents the first observation of an astrophysical muon neutrino flux in a model independent spectral measurement.

\begin{acknowledgements}

We acknowledge the support from the following agencies:
U.S. National Science Foundation-Office of Polar Programs,
U.S. National Science Foundation-Physics Division,
University of Wisconsin Alumni Research Foundation,
the Grid Laboratory Of Wisconsin (GLOW) grid infrastructure at the University of Wisconsin - Madison, the Open Science Grid (OSG) grid infrastructure;
U.S. Department of Energy, and National Energy Research Scientific Computing Center,
the Louisiana Optical Network Initiative (LONI) grid computing resources;
Natural Sciences and Engineering Research Council of Canada,
WestGrid and Compute/Calcul Canada;
Swedish Research Council,
Swedish Polar Research Secretariat,
Swedish National Infrastructure for Computing (SNIC),
and Knut and Alice Wallenberg Foundation, Sweden;
German Ministry for Education and Research (BMBF),
Deutsche Forschungsgemeinschaft (DFG),
Helmholtz Alliance for Astroparticle Physics (HAP),
Research Department of Plasmas with Complex Interactions (Bochum), Germany;
Fund for Scientific Research (FNRS-FWO),
FWO Odysseus programme,
Flanders Institute to encourage scientific and technological research in industry (IWT),
Belgian Federal Science Policy Office (Belspo);
University of Oxford, United Kingdom;
Marsden Fund, New Zealand;
Australian Research Council;
Japan Society for Promotion of Science (JSPS);
the Swiss National Science Foundation (SNSF), Switzerland;
National Research Foundation of Korea (NRF);
Villum Fonden, Danish National Research Foundation (DNRF), Denmark

\end{acknowledgements}

\nocite{CORSIKA}
\nocite{proposal}
\nocite{recopaper}
\nocite{Honda}
\nocite{Enberg}

\bibliographystyle{spphys}
\bibliography{refs}   

%


\end{document}